\documentclass[sigconf]{acmart}

\usepackage{subcaption}
\usepackage{tcolorbox}

\AtBeginDocument{%
  }

\setcopyright{acmlicensed}
\copyrightyear{2026}
\acmYear{2026}
\acmDOI{}

\acmConference[Tools for Thought @ CHI '26]
  {Tools for Thought @ CHI'26}
  {2026}
  {Barcelona, Spain}

\acmISBN{}

\begin{document}

\title{Exploring the Design of GenAI-Based Systems to Support Socially Shared Metacognition}


\author{Yihang Zhao}
\email{yihang.zhao@kcl.ac.uk}
\orcid{0009-0009-2436-8145}
\affiliation{%
  \institution{King's College London}
  \city{London}
  \country{United Kingdom}
}

\author{Wenxin Zhang}
\email{wenxin.3.zhang@kcl.ac.uk}
\orcid{0009-0007-7226-4928}
\affiliation{%
  \institution{King's College London}
  \city{London}
  \country{United Kingdom}
}

\author{Amy Rechkemmer}
\email{amy.rechkemmer@kcl.ac.uk}
\orcid{0000-0001-7572-751X}
\affiliation{%
  \institution{King's College London}
  \city{London}
  \country{United Kingdom}
}

\author{Albert Mero\~no Pe\~nuela}
\email{albert.merono@kcl.ac.uk}
\orcid{0000-0003-4646-5842}
\affiliation{%
  \institution{King's College London}
  \city{London}
  \country{United Kingdom}
}

\author{Elena Simperl}
\email{elena.simperl@kcl.ac.uk}
\orcid{0000-0003-1722-947X}
\affiliation{%
  \institution{King's College London}
  \city{London}
  \country{United Kingdom}
}
\affiliation{%
  \institution{Technical University of Munich}
  \city{Munich}
  \country{Germany}
}

\renewcommand{\shortauthors}{Zhao et al.}

\begin{abstract}

Socially shared metacognition (SSM) refers to the collective monitoring and regulation of joint cognitive processes in collaborative problem-solving, and is essential for effective knowledge work and learning. Generative AI (GenAI)-based systems offer new opportunities to support SSM, but emerging evidence suggests that poorly designed systems can encourage over-reliance on AI-generated explicit instruction and erode groups' capacity to develop autonomous regulatory processes. Group awareness tools (GATs) address this challenge through established design principles that make social and cognitive awareness information visible, highlight differences between group members to create cognitive conflict, and trigger autonomous elaboration and discussion, thereby implicitly guiding autonomous SSM emergence. This paper explores the design of GenAI-augmented GATs to support autonomous SSM in collaborative work and learning through an initial literature search, presenting preliminary design principles for discussion.

\end{abstract}

\ccsdesc[500]{Human-centered computing~Human computer interaction (HCI)}
\ccsdesc[500]{Computing methodologies~Artificial intelligence}

\keywords{generative AI, GenAI, large language models, LLMs, artificial intelligence, AI, cognition, metacognition, socially shared metacognition, SSM, co-regulation, self-regulation, autonomy, autonomous regulatory processes, explicit instruction, implicit guidance, group awareness, group awareness tools, GATs, computer-supported collaborative learning, CSCL, collaborative problem-solving, knowledge work, interaction design, user interface design, visualization}




\maketitle

\noindent\textbf{Author's Accepted Manuscript (AAM).} 
This is the Author's Accepted Manuscript version of the article: Zhao, Y., Zhang, W., Rechkemmer, A., Meroño-Peñuela, A., \& Simperl, E. (2026). Exploring the Design of GenAI-Based Systems to Support Socially Shared Metacognition. Accepted for publication in \textit{Tools for Thought @ CHI 2026}.

\section{Introduction}

Collaboration in modern knowledge work and learning environments has become increasingly complex. Team members may need to navigate distributed communication platforms, reconcile diverse disciplinary perspectives, synthesize information scattered across multiple sources, coordinate asynchronously across locations, and continuously adapt to evolving circumstances \cite{slade2023collaborative, an2025revealing}. Given these demands, the ability to collaborate effectively has become fundamental for both workers and learners \cite{ng2023teachers, jarvela2020bridging}.

Central to effective collaboration is socially shared metacognition (SSM), where group members collectively monitor and regulate their joint cognitive processes in collaborative problem-solving \cite{hadwin2010innovative, hurme2009socially}. Rather than each person independently tracking their own understanding, the entire group works together to assess collective knowledge, identify gaps, plan approaches, monitor progress, and adjust strategies, among other activities \cite{hadwin2011self, hadwin2017self}. When groups engage in SSM, they may achieve stronger decision-making, more effective problem-solving, and deeper shared knowledge construction, among other benefits \cite{kirschner2015awareness, volet2009self, iiskala2011socially}.

Yet SSM rarely emerges spontaneously \cite{tsai2018exploring, volet2009high}, creating needs for intervention \cite{silva2023fostering, panadero2015socially}. Thanks to their understanding and generative capacity \cite{zhao2026ontoscope, hu2025designing, zhao2025leveraging, zhaosurvey, zhao2024improving, zhao2024using}, GenAI-based systems present new possibilities for supporting SSM through, for example, automated monitoring of group processes and timely feedback provision \cite{george2023managing, song2021survey}. However, poorly designed GenAI-based systems may encourage over-reliance on AI-generated explicit instruction and erode groups' capacity to develop autonomous regulatory processes \cite{liu2023incorporating, yang2025analysing, kim2025socially, bauer2025looking}.

GATs address this challenge through established design principles that make social and cognitive awareness information visible, highlight differences between group members to potentially create cognitive conflict, and may trigger autonomous elaboration and discussion, thereby implicitly guiding autonomous SSM emergence \cite{sangin2011facilitating, schnaubert2020combining}. GATs externalize information about group functioning through visualizations such as participation patterns, knowledge distributions, and progress indicators \cite{jarvela2015enhancing, lin2018effects}. While GATs interpret collaboration data to generate awareness information, they present this as observable patterns for groups to examine rather than prescriptive actions to follow \cite{engelmann2009knowledge, sangin2011facilitating}, potentially preserving groups' autonomy over monitoring, evaluation, and planning activities \cite{sangin2009peer}.

This paper explores the design of GenAI-augmented GATs to support autonomous SSM in collaborative work and learning contexts. Through an initial literature search of existing GAT systems, we address three research questions:

\begin{itemize}
    \item \textbf{RQ1:} What group awareness information can GenAI generate?
    \item \textbf{RQ2:} How can user interfaces present GenAI-generated group awareness information?
    \item \textbf{RQ3:} How can interaction techniques enable exploration of GenAI-generated group awareness information?
\end{itemize}

Our contribution addresses the workshop's ``TfT Strategies: Design and Usage'' theme by offering preliminary design principles for GenAI-based systems that protect human cognitive processes while leveraging AI capabilities where they add value.

\section{Background and Related Work}

SSM encompasses three regulatory processes through which groups collectively control their collaborative work \cite{iiskala2011socially, iiskala2015socially}. In shared planning, groups jointly determine task approaches, role distributions, and resource allocation, among other activities \cite{hadwin2017self}. During shared monitoring, groups track ongoing progress, contribution patterns, and emerging outcomes \cite{biasutti2018group}. Through shared evaluation, groups assess results and processes against criteria such as initial goals and quality standards \cite{biasutti2018group, hadwin2017self}. These processes may operate cyclically: evaluation insights can inform subsequent planning, which shapes what groups monitor, leading to further evaluation.

Supporting these regulatory processes requires groups to develop group awareness, understood as knowledge of their collective state including how the collaboration functions and how knowledge distributes among members \cite{bodemer2011group, engelmann2009evoking, janssen2011group}. Group awareness may encompass cognitive dimensions (e.g., what members know and understand about tasks) and social dimensions (e.g., how members participate and interact) \cite{janssen2013coordinated, bodemer2011tacit, janssen2007visualization}.

GATs support SSM by externalizing group awareness information through visual representations \cite{jarvela2015enhancing, lin2018effects}. For example, cognitive GATs may display comprehension levels through bar charts or knowledge structures through concept maps \cite{engelmann2011fostering, sangin2011facilitating}. Social GATs may visualize participation patterns through contribution counts or interaction quality through discourse analysis \cite{jongsawat2009empirical, janssen2007visualization}. Critically, GATs employ a distinct support mechanism: rather than providing explicit instruction on what groups should do, they present awareness information as observable patterns that groups interpret themselves \cite{engelmann2009knowledge, sangin2011facilitating}, potentially creating cognitive conflict through visible discrepancies such as uneven participation, knowledge gaps, or performance shortfalls, which may trigger autonomous elaboration and discussion \cite{kirschner2015awareness, sangin2011facilitating}. This approach has the potential to preserve monitoring, evaluation, and planning within group control rather than delegating these regulatory processes to the system \cite{sangin2009peer}.

GenAI introduces new possibilities for augmenting GATs \cite{de2024assessing, claggett2025relational}. Traditional GATs typically display raw data or simple visualizations derived from structured collaboration data \cite{zamecnik2022team, chang2024effects}. GenAI may analyze unstructured collaboration artifacts such as discussion transcripts or document revisions \cite{breideband2025feasibility, suraworachet2025university}, potentially enabling richer awareness information about nuanced collaboration aspects, such as whether members build on each other's ideas, who demonstrates deeper understanding, or which misconceptions are developing \cite{suraworachet2025university, hu2025conversational}. However, integrating GenAI while preserving GATs' core principle of implicit guidance rather than explicit instruction requires careful design consideration.

\section{Methodology}
To develop preliminary design principles for GenAI-augmented GATs, we conducted an initial literature search of existing GAT systems that support SSM in collaborative contexts. We searched three digital libraries, including ACM Digital Library, IEEE Xplore, and Scopus, using search terms combining concepts such as ``group awareness tools'', ``socially shared metacognition'', ``shared regulation'', ``collaborative metacognition'', and ``awareness visualizations'', among others. We did not include explicit terms for GenAI, since all GATs, whether GenAI-based or not, may serve as potential sources for informing GenAI-augmented system design.

We applied lightweight inclusion and exclusion criteria suited to the exploratory scope of this workshop paper. We included peer-reviewed papers describing systems with UI designs or interaction techniques for group awareness externalization that discuss their role in facilitating SSM or related constructs. We excluded secondary studies, purely theoretical papers, and systems relying on non-traditional modalities such as augmented or virtual reality, as these warrant separate investigation. To complement the initial database search, we conducted a round of backward snowballing on included papers to identify additional relevant systems that might otherwise have been missed.

The resulting set of papers was analyzed thematically to identify patterns in how GATs generate, present, and support exploration of group awareness information. These patterns informed the three preliminary design principles discussed in Section~\ref{sec:findings}. We acknowledge that this initial search is not exhaustive, and a more rigorous systematic review would be needed to validate and refine these principles in future work.

\section{Initial Findings}\label{sec:findings}

\subsection{Generating Group Awareness Information}
GenAI's distinguishing capability may lie in interpreting unstructured collaboration content. Traditional GATs typically employ rule-based approaches that apply predefined logic to structured data \cite{janssen2013coordinated, bodemer2011tacit}, and may excel at tasks such as counting message frequencies, calculating participation ratios, or aggregating self-reported understanding levels. However, they may struggle with awareness information requiring interpretation of unstructured content, such as assessing whether members demonstrate shared understanding, identifying reasoning quality in discussions, or detecting emerging conflicts from language cues \cite{breideband2025feasibility}.

GenAI may analyze unstructured collaboration artifacts such as discussion transcripts, document revisions, or annotations to infer cognitive and social states that resist formalization into explicit rules \cite{suraworachet2025university, breideband2025feasibility}. For example, rather than merely counting contributions, GenAI might examine what members wrote to assess whether explanations demonstrate conceptual understanding, or whether members build on versus talk past each other.

However, GenAI's value may vary substantially across awareness information types. For awareness information requiring semantic understanding of unstructured content, GenAI may offer clear advantages, such as inferring collaboratively defined norms, shared understanding, communication quality, or reasoning processes. For awareness information derivable through straightforward computation of structured data, traditional approaches may remain more suitable, including individual contribution quantities, participation patterns over time, and activity completion status \cite{zamecnik2022team, chang2024effects}. Between these extremes lies awareness information that could potentially benefit from GenAI's semantic capabilities but may also be adequately generated traditionally, depending on available data structures. For example, quality of collaborative processes could be analyzed through GenAI discourse analysis or assessed through predefined rubrics; comprehension development could be tracked through GenAI analysis of evolving contributions or through periodic structured assessments; and participation distribution could potentially benefit from GenAI's ability to distinguish substantive from superficial contributions, though basic quantitative metrics may remain useful for identifying gross imbalances.

\begin{tcolorbox}[colback=blue!5!white, colframe=blue!40!white, arc=4pt, boxrule=0.6pt]
\textbf{Design Principle 1:} \textit{Employ hybrid architectures combining rule-based systems for quantitative metrics requiring computational precision with GenAI for qualitative assessments requiring semantic understanding of unstructured collaboration content.}
\end{tcolorbox}

\subsection{Presenting Awareness Information}
A critical challenge emerges when presenting GenAI-generated awareness information: groups may perceive GenAI's semantic assessments as definitive judgments rather than provisional interpretations \cite{bauer2025looking, yan2025distinguishing}, potentially undermining SSM by discouraging independent critical evaluation. We propose a core design strategy: GenAI-generated semantic interpretations should appear as \textit{secondary} visual encodings that augment rather than replace quantitative representations. By encoding GenAI assessments through visual properties such as color saturation, background shading, or overlay indicators, while preserving traditional quantitative metrics as primary visual channels, groups can perceive both objective quantities (what occurred) and semantic qualities (what it might mean) simultaneously \cite{jarvela2023human, edwards2025human, zheng2025cognitive}. Critically, this dual encoding may create cognitive conflict: when groups observe discrepancies between their primary quantitative view and the secondary GenAI-generated interpretation, this psychological discomfort may trigger autonomous elaboration and discussion, prompting groups to interrogate the difference rather than passively accept either representation.

\begin{figure}[h]
    \centering
    \begin{subfigure}[t]{0.48\linewidth}
        \centering
        \includegraphics[width=\linewidth]{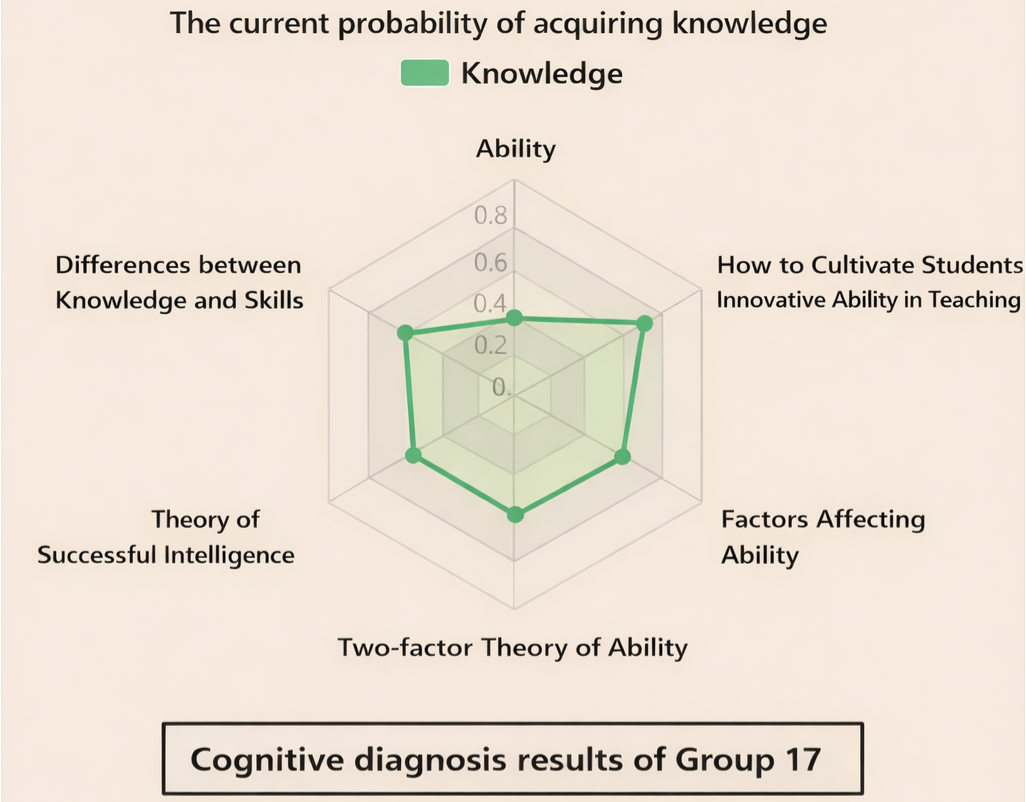}
        \caption{Without GenAI integration (reproduced from \cite{zheng2025novel}, translated into English with minor visual revisions).}
        \label{fig:RQ23}
    \end{subfigure}
    \hfill
    \begin{subfigure}[t]{0.48\linewidth}
        \centering
        \includegraphics[width=\linewidth]{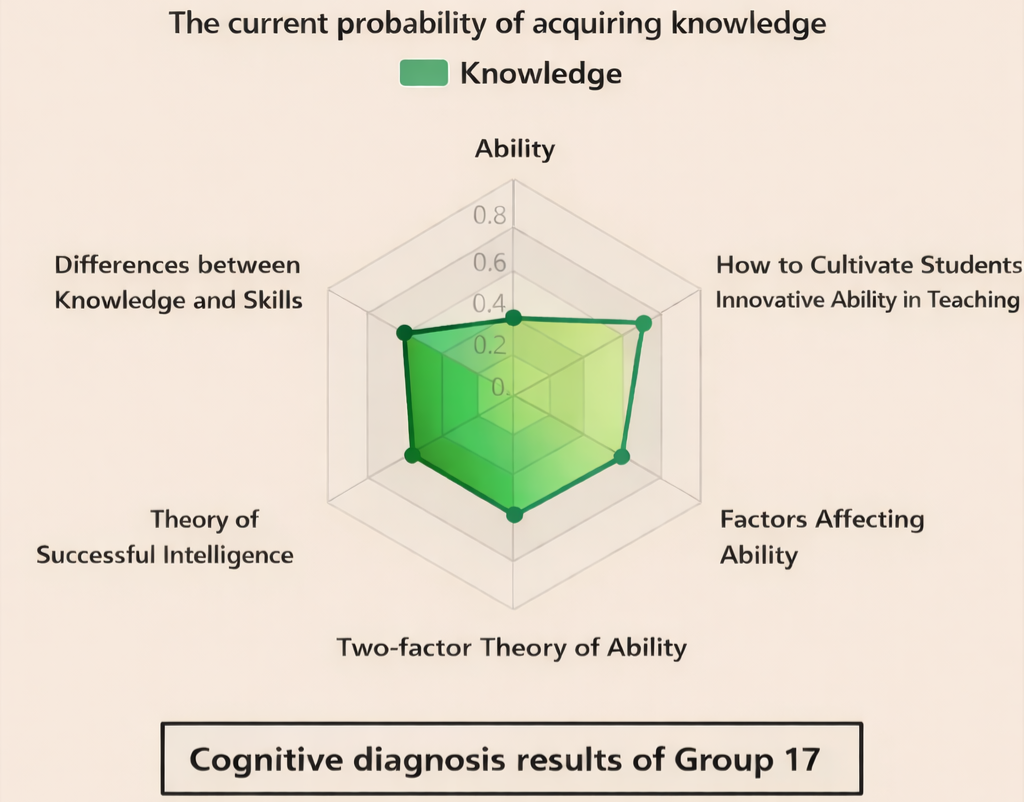}
        \caption{With GenAI integration.}
        \label{fig:RQ23G}
    \end{subfigure}
    \caption{Comparison of radar and spider chart UIs without and with GenAI integration.}
    \Description{Figure shows two radar charts side by side. The left chart displays a traditional radar chart with a polygon shape representing group self-reported knowledge levels across multiple knowledge domains. The right chart shows a GenAI-augmented version where the same polygon is preserved but axis segments have varying background color intensities, with darker backgrounds indicating alignment between self-reported and demonstrated understanding, and lighter backgrounds indicating differences.}
    \label{fig:radar-chart-comparison}
\end{figure}

Consider radar charts showing knowledge levels across domains (Figure~\ref{fig:radar-chart-comparison}) \cite{zheng2025novel}. Rather than mapping GenAI analysis directly onto the shape of the polygon, which would position AI assessment as the sole source of information, a GenAI-augmented design might maintain the polygon representing group self-reported understanding while showing GenAI's independent analysis of actual discussions through background color intensity on axis segments. Darker backgrounds may indicate alignment between what groups reported and what they demonstrated in discussions, while lighter backgrounds may indicate differences. This juxtaposition may create cognitive conflict: groups observing a light background on an axis where they reported high understanding may be prompted to question their self-assessment, examine whether their discussions genuinely reflected that understanding, and decide whether additional work is needed or whether the GenAI analysis missed important context. Rather than instructing groups on what to do, this design implicitly guides autonomous monitoring and evaluation by making the tension between self-perception and demonstrated performance visible and open to group interpretation.

\begin{tcolorbox}[colback=blue!5!white, colframe=blue!40!white, arc=4pt, boxrule=0.6pt]
\textbf{Design Principle 2:} \textit{Present GenAI-generated semantic interpretations as secondary visual encodings such as color saturation, background intensity, or overlay indicators, that augment primary quantitative representations, creating cognitive conflict through visible discrepancies that may trigger autonomous elaboration and discussion while preserving groups' interpretive autonomy.}
\end{tcolorbox}

\subsection{Enabling Exploration Through Interaction}
Beyond generating and presenting awareness information, GATs may provide interaction techniques enabling groups to explore visualized information. Such exploration may contribute to autonomous SSM by enabling groups to verify, question, and critically evaluate AI interpretations rather than passively accepting them, particularly when groups observe discrepancies between GenAI's visualized interpretations and their own experience.

\begin{figure}[h]
    \centering
    \includegraphics[width=0.96\linewidth]{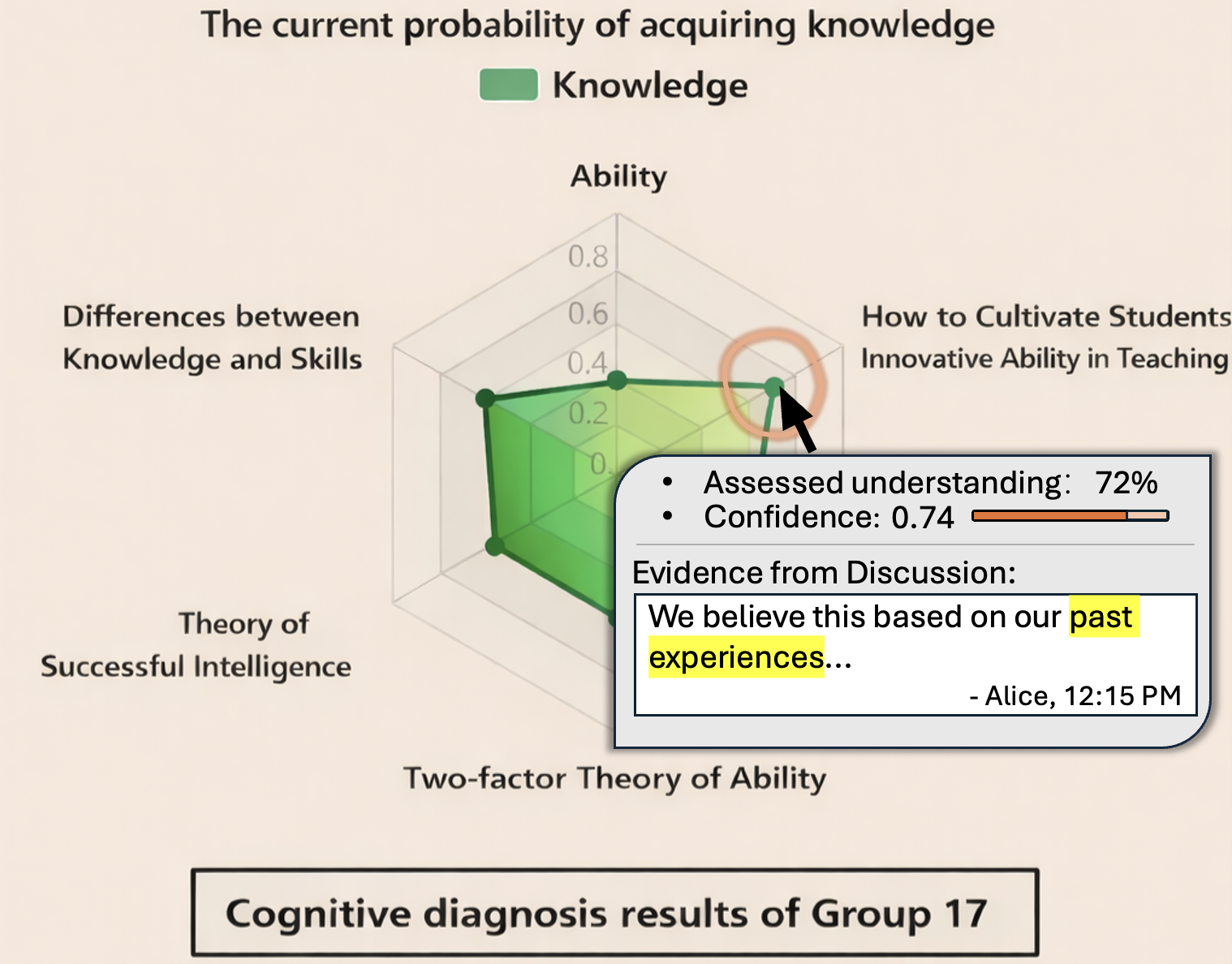}
    \caption{Hover-for-details interaction with radar charts.}
    \Description{A GenAI-augmented radar chart with background color intensities on axis segments. A pop-up window appears when hovering over an axis, displaying GenAI's assessed understanding level, confidence score, and example discussion quotes as evidence.}
    \label{fig:interaction-hover}
\end{figure}

Hover-for-details interactions involve moving the cursor over visual elements to reveal pop-ups \cite{dulger2025designing, hu2025conversational}. Applied to radar charts (Figure~\ref{fig:interaction-hover}), when groups observe axes with light background intensity indicating differences between self-reported and AI-assessed understanding, hovering over those axes might trigger pop-ups displaying several layers of information. For example, the pop-up might show GenAI's assessed understanding level for that specific knowledge domain, making explicit what the background color intensity represents. It may also display a confidence score indicating how certain GenAI is about its assessment, potentially helping groups calibrate their trust in the interpretation. Additionally, it might present example discussion quotes that GenAI identified as evidence for its assessment, allowing groups to examine the actual statements that informed the AI's judgment. Together, this multi-layered information may enable groups to investigate why GenAI assessed understanding differently than the group self-reported, revealing both the system's reasoning and its uncertainty. Groups might then verify whether the quoted evidence aligns with their own interpretation, and decide whether the AI assessment warrants revising their self-perception or whether important context was missed.

Additional interaction techniques may further support exploration. Click-to-access interactions involve clicking visual elements to reveal underlying evidence \cite{chen2021spiral, hu2025conversational, yin2025association}, such as opening panels showing full discussion excerpts with highlighting that matches quality assessments. Selection and highlighting interactions involve clicking items to mark them for comparison \cite{farrokhnia2025improving}, such as selecting multiple group members to view side-by-side statistics and example contributions with quality indicators. Together, these interaction techniques may help transform cognitive conflict created by the dual visual encoding into productive autonomous regulatory activity, as groups move from noticing discrepancies to actively investigating, evaluating, and deciding how to respond.

\begin{tcolorbox}[colback=blue!5!white, colframe=blue!40!white, arc=4pt, boxrule=0.6pt]
\textbf{Design Principle 3:} \textit{Provide interaction techniques such as click-to-access underlying evidence, hover-for-details on assessments, and selection for comparison, that enable groups to investigate evidence underlying AI interpretations, compare AI assessments with their own experiences, and critically evaluate whether assessments are valid before deciding whether to act on them.}
\end{tcolorbox}

\section{Discussion}
The design principles proposed in this paper operationalize GenAI as a tool for thought that protects rather than erodes autonomous regulatory processes in collaborative work and learning. By presenting AI-generated interpretations as secondary visual encodings that augment quantitative representations, these designs avoid positioning GenAI as an authoritative evaluator. Instead, GenAI may perform pattern detection that groups would struggle to accomplish manually with large volumes of unstructured collaboration data, while preserving regulatory decisions such as what patterns mean, whether they indicate problems, and how to respond, under group control.

The proposed principles may transfer across educational and workplace contexts because they address fundamental aspects of how groups develop awareness and engage in regulatory processes rather than domain-specific task characteristics. While most existing GAT research originates from educational settings \cite{jarvela2015enhancing, hadwin2017self}, workplace collaborations involving distributed expertise and collective decision-making may equally require effective SSM, though contextual adaptations are likely needed.

Additionally, a critical assumption underlying these principles is that groups will engage productively with GenAI-generated awareness information rather than finding it distracting or overwhelming during active collaboration. Several design considerations may help mitigate these risks. Awareness information might be surfaced at natural transition points rather than updating continuously during active task work. Progressive disclosure may also help: systems might initially surface only primary quantitative representations, with secondary GenAI encodings revealed on demand through interaction techniques. Finally, the salience of secondary visual encodings may need careful calibration, as encodings that are too prominent may disrupt task attention while encodings that are too subtle may fail to trigger cognitive conflict. These considerations suggest a direction for future design exploration.

We hope to discuss these considerations in this workshop, and the expertise we can potentially bring spans machine learning, AI-based systems for individual and collaborative knowledge and creative work, interaction design with AI, and cognitive and metacognitive augmentation \cite{zhao2026ontoscope,zhao2025leveraging,zhaosurvey,zhao2024improving,zhao2024using,zhao2026designguidanceaddressingoverreliance,zhao2024user,hu2025designing,lisena2026data,simperl2025automatic,simperl2025introducing,zhang2025trustworthy,zhao2021discriminative,luo2021survey}.

\section{Limitations}
The proposed principles represent conceptually derived guidelines grounded in existing GAT patterns but lack empirical validation regarding their effectiveness in supporting autonomous SSM, their potential to create cognitive overload, or their capacity to engage groups productively in real collaborative settings. Future work should validate these principles through controlled studies and field deployments across diverse contexts.

\section{Conclusion}
This paper explored how GenAI-augmented GATs might support autonomous SSM in collaborative work and learning contexts. We proposed three preliminary design principles: employing hybrid architectures that deploy GenAI selectively for qualitative awareness information generation; presenting GenAI-generated awareness information as secondary visual encodings that may create cognitive conflict to trigger autonomous elaboration and discussion; and providing interaction techniques enabling groups to critically explore and evaluate AI interpretations. Together, these principles operationalize GenAI as a tool for thought that may augment rather than replace human cognitive processes, supporting implicit guidance rather than explicit instruction, thereby preserving rather than eroding groups' autonomous regulatory processes. Realizing this potential, however, requires careful attention to when and how awareness information is surfaced to engage rather than distract or overwhelm groups during active collaboration.


\begin{acks}
We thank Advait Sarkar from Microsoft Research Cambridge for feedback. We also acknowledge funding support, including the Engineering and Physical Sciences Research Council [grant number EP/Y009800/1], funded through Responsible AI UK (KP0011); SIEMENS AG; and the Institute for Advanced Study, Technical University of Munich, Germany.
\end{acks}

\bibliographystyle{ACM-Reference-Format}
\bibliography{tft}

@article{iiskala2011socially,
  title={Socially shared metacognition of dyads of pupils in collaborative mathematical problem-solving processes},
  author={Iiskala, Tuike and Vauras, Marja and Lehtinen, Erno and Salonen, Pekka},
  journal={Learning and instruction},
  volume={21},
  number={3},
  pages={379--393},
  year={2011},
  publisher={Elsevier}
}

@article{panadero2015socially,
  title={Socially shared regulation of learning: A review},
  author={Panadero, Ernesto and J{\"a}rvel{\"a}, Sanna},
  journal={European psychologist},
  year={2015},
  publisher={Hogrefe Publishing}
}

@article{volet2009self,
  title={Self-and social regulation in learning contexts: An integrative perspective},
  author={Volet, Simone and Vauras, Marja and Salonen, Pekka},
  journal={Educational psychologist},
  volume={44},
  number={4},
  pages={215--226},
  year={2009},
  publisher={Taylor \& Francis}
}

@article{ng2023teachers,
  title={Teachers’ AI digital competencies and twenty-first century skills in the post-pandemic world},
  author={Ng, Davy Tsz Kit and Leung, Jac Ka Lok and Su, Jiahong and Ng, Ross Chi Wui and Chu, Samuel Kai Wah},
  journal={Educational technology research and development},
  volume={71},
  number={1},
  pages={137--161},
  year={2023},
  publisher={Springer}
}

@article{an2025revealing,
  title={Revealing the interplay of cognitive, meta-cognitive, and social processes in university students’ collaborative problem solving: A three-stage analytical framework},
  author={An, Shuowen and Zhang, Si and Cai, Zhihui and Pan, Wei and Li, Mingwei and Tong, Mingwen},
  journal={International Journal of Computer-Supported Collaborative Learning},
  volume={20},
  number={1},
  pages={9--39},
  year={2025},
  publisher={Springer}
}

@article{jarvela2020bridging,
  title={Bridging learning sciences, machine learning and affective computing for understanding cognition and affect in collaborative learning},
  author={J{\"a}rvel{\"a}, Sanna and Ga{\v{s}}evi{\'c}, Dragan and Sepp{\"a}nen, Tapio and Pechenizkiy, Mykola and Kirschner, Paul A},
  journal={British Journal of Educational Technology},
  volume={51},
  number={6},
  pages={2391--2406},
  year={2020},
  publisher={Wiley Online Library}
}

@article{hurme2009socially,
  title={Socially shared metacognition of pre-service primary teachers in a computer-supported mathematics course and their feelings of task difficulty: A case study},
  author={Hurme, Tarja-Riitta and Merenluoto, Kaarina and J{\"a}rvel{\"a}, Sanna},
  journal={Educational Research and Evaluation},
  volume={15},
  number={5},
  pages={503--524},
  year={2009},
  publisher={Taylor \& Francis}
}

@article{biasutti2018group,
  title={Group metacognition in online collaborative learning: Validity and reliability of the group metacognition scale (GMS)},
  author={Biasutti, Michele and Frate, Sara},
  journal={Educational Technology Research and Development},
  volume={66},
  number={6},
  pages={1321--1338},
  year={2018},
  publisher={Springer}
}

@article{iiskala2015socially,
  title={Socially shared metacognitive regulation in asynchronous CSCL in science: Functions, evolution and participation.},
  author={Iiskala, Tuike and Volet, Simone and Lehtinen, Erno and Vauras, Marja},
  journal={Frontline Learning Research},
  volume={3},
  number={1},
  pages={78--111},
  year={2015},
  publisher={ERIC}
}

@article{hadwin2010innovative,
  title={Innovative ways for using gStudy to orchestrate and research social aspects of self-regulated learning},
  author={Hadwin, Allyson F and Oshige, Mika and Gress, Carmen LZ and Winne, Philip H},
  journal={Computers in Human behavior},
  volume={26},
  number={5},
  pages={794--805},
  year={2010},
  publisher={Elsevier}
}

@article{jarvela2015enhancing,
  title={Enhancing socially shared regulation in collaborative learning groups: Designing for CSCL regulation tools},
  author={J{\"a}rvel{\"a}, Sanna and Kirschner, Paul A and Panadero, Ernesto and Malmberg, Jonna and Phielix, Chris and Jaspers, Jos and Koivuniemi, Marika and J{\"a}rvenoja, Hanna},
  journal={Educational Technology Research and Development},
  volume={63},
  number={1},
  pages={125--142},
  year={2015},
  publisher={Springer}
}

@article{volet2009high,
  title={High-level co-regulation in collaborative learning: How does it emerge and how is it sustained?},
  author={Volet, Simone and Summers, Mark and Thurman, Joanne},
  journal={Learning and Instruction},
  volume={19},
  number={2},
  pages={128--143},
  year={2009},
  publisher={Elsevier}
}

@article{tsai2018exploring,
  title={Exploring the effects of web-mediated socially-shared regulation of learning and experience-based learning on improving students’ learning},
  author={Tsai, Chia-Wen and Shen, Pei-Di and Chiang, I-Chun and Chen, Wen-Yu and Chen, Yi-Fen},
  journal={Interactive Learning Environments},
  volume={26},
  number={6},
  pages={815--826},
  year={2018},
  publisher={Taylor \& Francis}
}

@article{edwards2025human,
  title={Human-AI collaboration: Designing artificial agents to facilitate socially shared regulation among learners},
  author={Edwards, Justin and Nguyen, Andy and L{\"a}ms{\"a}, Joni and Sobocinski, Marta and Whitehead, Ridwan and Dang, Belle and Roberts, Anni-Sofia and J{\"a}rvel{\"a}, Sanna},
  journal={British Journal of Educational Technology},
  volume={56},
  number={2},
  pages={712--733},
  year={2025},
  publisher={Wiley Online Library}
}

@article{jarvela2023human,
  title={Human and artificial intelligence collaboration for socially shared regulation in learning},
  author={J{\"a}rvel{\"a}, Sanna and Nguyen, Andy and Hadwin, Allyson},
  journal={British Journal of Educational Technology},
  volume={54},
  number={5},
  pages={1057--1076},
  year={2023},
  publisher={Wiley Online Library}
}

@article{engelmann2011fostering,
  title={Fostering sharing of unshared knowledge by having access to the collaborators’ meta-knowledge structures},
  author={Engelmann, Tanja and Hesse, Friedrich W},
  journal={Computers in Human Behavior},
  volume={27},
  number={6},
  pages={2078--2087},
  year={2011},
  publisher={Elsevier}
}

@article{kirschner2015awareness,
  title={Awareness of cognitive and social behaviour in a CSCL environment},
  author={Kirschner, Paul A and Kreijns, Karel and Phielix, Chris and Fransen, Jos},
  journal={Journal of Computer Assisted Learning},
  volume={31},
  number={1},
  pages={59--77},
  year={2015},
  publisher={Wiley Online Library}
}

@article{bodemer2011tacit,
  title={Tacit guidance for collaborative multimedia learning},
  author={Bodemer, Daniel},
  journal={Computers in Human Behavior},
  volume={27},
  number={3},
  pages={1079--1086},
  year={2011},
  publisher={Elsevier}
}

@article{janssen2011group,
  title={Group awareness tools: It’s what you do with it that matters},
  author={Janssen, Jeroen and Erkens, Gijsbert and Kirschner, Paul A},
  journal={Computers in human behavior},
  volume={27},
  number={3},
  pages={1046--1058},
  year={2011},
  publisher={Elsevier}
}

@article{janssen2007visualization,
  title={Visualization of agreement and discussion processes during computer-supported collaborative learning},
  author={Janssen, Jeroen and Erkens, Gijsbert and Kanselaar, Gellof},
  journal={Computers in human behavior},
  volume={23},
  number={3},
  pages={1105--1125},
  year={2007},
  publisher={Elsevier}
}

@article{janssen2013coordinated,
  title={Coordinated computer-supported collaborative learning: Awareness and awareness tools},
  author={Janssen, Jeroen and Bodemer, Daniel},
  journal={Educational psychologist},
  volume={48},
  number={1},
  pages={40--55},
  year={2013},
  publisher={Taylor \& Francis}
}

@article{bauer2025looking,
  title={Looking beyond the hype: Understanding the effects of AI on learning},
  author={Bauer, Elisabeth and Greiff, Samuel and Graesser, Arthur C and Scheiter, Katharina and Sailer, Michael},
  journal={Educational Psychology Review},
  volume={37},
  number={2},
  pages={45},
  year={2025},
  publisher={Springer}
}

@article{yang2025analysing,
  title={Analysing nontraditional students' ChatGPT interaction, engagement, self-efficacy and performance: A mixed-methods approach},
  author={Yang, Mohan and Jiang, Shiyan and Li, Belle and Herman, Kristin and Luo, Tian and Moots, Shanan Chappell and Lovett, Nolan},
  journal={British Journal of Educational Technology},
  year={2025},
  publisher={Wiley Online Library}
}

@article{kim2025socially,
  title={Socially shared regulation of learning and artificial intelligence: Opportunities to support socially shared regulation},
  author={Kim, Jinhee and Detrick, Rita and Yu, Seongryeong and Song, Yukyeong and Bol, Linda and Li, Na},
  journal={Education and Information Technologies},
  pages={1--39},
  year={2025},
  publisher={Springer}
}

@article{yan2025distinguishing,
  title={Distinguishing performance gains from learning when using generative AI},
  author={Yan, Lixiang and Greiff, Samuel and Lodge, Jason M and Ga{\v{s}}evi{\'c}, Dragan},
  journal={Nature Reviews Psychology},
  pages={1--2},
  year={2025},
  publisher={Nature Publishing Group US New York}
}

@phdthesis{sangin2009peer,
  title={Peer knowledge modeling in computer supported collaborative learning},
  author={Sangin, Mirweis},
  year={2009},
  school={Verlag nicht ermittelbar}
}

@article{sangin2011facilitating,
  title={Facilitating peer knowledge modeling: Effects of a knowledge awareness tool on collaborative learning outcomes and processes},
  author={Sangin, Mirweis and Molinari, Ga{\"e}lle and N{\"u}ssli, Marc-Antoine and Dillenbourg, Pierre},
  journal={Computers in human behavior},
  volume={27},
  number={3},
  pages={1059--1067},
  year={2011},
  publisher={Elsevier}
}

@article{schnaubert2020combining,
  title={Combining scripts, group awareness tools and self-regulated learning--theoretical implications and practical implementations},
  author={Schnaubert, Lenka and Vogel, Freydis and Bodemer, Daniel and Fischer, Frank and Radkowitsch, Anika and Schmidmaier, Ralf and Fischer, Martin R and Tsovaltzi, Dimitra and Puhl, Thomas and Azevedo, Roger},
  year={2020},
  publisher={International Society of the Learning Sciences (ISLS)}
}

@article{engelmann2009knowledge,
  title={Knowledge awareness in CSCL: A psychological perspective},
  author={Engelmann, Tanja and Dehler, Jessica and Bodemer, Daniel and Buder, J{\"u}rgen},
  journal={Computers in Human Behavior},
  volume={25},
  number={4},
  pages={949--960},
  year={2009},
  publisher={Elsevier}
}

@article{hadwin2011self,
  title={Self-regulation, coregulation, and socially shared regulation: Exploring perspectives of social in self-regulated learning theory},
  author={Hadwin, Allyson and Oshige, Mika},
  journal={Teachers College Record},
  volume={113},
  number={2},
  pages={240--264},
  year={2011},
  publisher={SAGE Publications Sage CA: Los Angeles, CA}
}

@incollection{hadwin2017self,
  title={Self-regulation, co-regulation, and shared regulation in collaborative learning environments},
  author={Hadwin, Allyson and J{\"a}rvel{\"a}, Sanna and Miller, Mariel},
  booktitle={Handbook of self-regulation of learning and performance},
  pages={83--106},
  year={2017},
  publisher={Routledge}
}

@article{lin2018effects,
  title={Effects of an online team project-based learning environment with group awareness and peer evaluation on socially shared regulation of learning and self-regulated learning},
  author={Lin, Jian-Wei},
  journal={Behaviour \& Information Technology},
  volume={37},
  number={5},
  pages={445--461},
  year={2018},
  publisher={Taylor \& Francis}
}

@article{silva2023fostering,
  title={Fostering regulatory processes using computational scaffolding},
  author={Silva, Leonardo and Mendes, Ant{\'o}nio and Gomes, Anabela and Fortes, Gabriel},
  journal={International Journal of Computer-Supported Collaborative Learning},
  volume={18},
  number={1},
  pages={67--100},
  year={2023},
  publisher={Springer}
}

@article{bodemer2011group,
  title={Group awareness in CSCL environments},
  author={Bodemer, Daniel and Dehler, Jessica},
  journal={Computers in Human Behavior},
  volume={27},
  number={3},
  pages={1043--1045},
  year={2011},
  publisher={Elsevier}
}

@article{engelmann2009evoking,
  title={Evoking knowledge and information awareness for enhancing computer-supported collaborative problem solving},
  author={Engelmann, Tanja and Tergan, Sigmar-Olaf and Hesse, Friedrich W},
  journal={The Journal of Experimental Education},
  volume={78},
  number={2},
  pages={268--290},
  year={2009},
  publisher={Taylor \& Francis}
}

@inproceedings{jongsawat2009empirical,
  title={An empirical study of group awareness information in web-based group decision support system in a field test setting},
  author={Jongsawat, Nipat and Premchaiswadi, Wichian},
  booktitle={2009 7th International Conference on ICT and Knowledge Engineering},
  pages={15--23},
  year={2009},
  organization={IEEE}
}

@inproceedings{de2024assessing,
  title={Assessing Cognitive and Social Awareness among Group Members in AI-assisted Collaboration},
  author={de Jong, Sander and Wester, Joel and Schrills, Tim and S. Secher, Kristina and F. Griggio, Carla and van Berkel, Niels},
  booktitle={Proceedings of the International Conference on Mobile and Ubiquitous Multimedia},
  pages={338--350},
  year={2024}
}

@inproceedings{claggett2025relational,
  title={Relational ai: Facilitating intergroup cooperation with socially aware conversational support},
  author={Claggett, Elijah L and Kraut, Robert E and Shirado, Hirokazu},
  booktitle={Proceedings of the 2025 CHI Conference on Human Factors in Computing Systems},
  pages={1--22},
  year={2025}
}

@article{zheng2025cognitive,
  title={Cognitive Echo: Enhancing think-aloud protocols with LLM-based simulated students},
  author={Zheng, Longwei and He, Anna and Qi, Changyong and Zhang, Haomin and Gu, Xiaoqing},
  journal={British Journal of Educational Technology},
  year={2025},
  publisher={Wiley Online Library}
}

@article{yin2025association,
  title={The association between groups' interactions with the Visual-GenAI learning analytics feedback and student engagement in CSCL},
  author={Yin, Xinghan and Ye, Junmin and Yu, Shuang and Li, Honghui and Liu, Qingtang and Zhao, Gang},
  journal={Computers \& Education},
  pages={105434},
  year={2025},
  publisher={Elsevier}
}

@article{chen2021spiral,
  title={The spiral model of collaborative knowledge improvement: An exploratory study of a networked collaborative classroom},
  author={Chen, Wenli and Tan, Jesmine SH and Pi, Zhongling},
  journal={International Journal of Computer-Supported Collaborative Learning},
  volume={16},
  number={1},
  pages={7--35},
  year={2021},
  publisher={Springer}
}

@article{zamecnik2022team,
  title={Team interactions with learning analytics dashboards},
  author={Zamecnik, Andrew and Kovanovi{\'c}, Vitomir and Grossmann, Georg and Joksimovi{\'c}, Sre{\'c}ko and Jolliffe, Gabrielle and Gibson, David and Pardo, Abelardo},
  journal={Computers \& Education},
  volume={185},
  pages={104514},
  year={2022},
  publisher={Elsevier}
}

@article{chang2024effects,
  title={Effects of a peer assessment-based scrum project learning system on computer programming’s learning motivation, collaboration, communication, critical thinking, and cognitive load},
  author={Chang, Shao-Chen and Wongwatkit, Charoenchai},
  journal={Education and Information Technologies},
  volume={29},
  number={6},
  pages={7105--7128},
  year={2024},
  publisher={Springer}
}

@article{farrokhnia2025improving,
  title={Improving hybrid brainstorming outcomes with computer-supported scaffolds: Prompts and cognitive group awareness},
  author={Farrokhnia, Mohammadreza and Noroozi, Omid and Baggen, Yvette and Biemans, Harm and Weinberger, Armin},
  journal={Computers \& Education},
  volume={227},
  pages={105229},
  year={2025},
  publisher={Elsevier}
}

@article{dulger2025designing,
  title={Designing a classroom-level teacher dashboard to foster primary school teachers’ direct instruction of self-regulated learning strategies},
  author={D{\"u}lger, Melis and van Leeuwen, Anouschka and Janssen, Jeroen and Kester, Liesbeth},
  journal={Education and Information Technologies},
  pages={1--35},
  year={2025},
  publisher={Springer}
}

@article{hu2025conversational,
  title={A conversational agent based on contingent teaching model to support collaborative learning activities: impacts on students’ learning performance, self-efficacy and perceptions: W. Hu et al.},
  author={Hu, Wanqing and Gong, Rushi and Wu, Sisi and Li, Yanyan},
  journal={Educational technology research and development},
  pages={1--32},
  year={2025},
  publisher={Springer}
}

@article{suraworachet2025university,
  title={University Students' Perceptions of a Multimodal AI System for Real-World Collaboration Analytics: Lessons Learned From a Case Study},
  author={Suraworachet, Wannapon and Zhou, Qi and Cukurova, Mutlu},
  journal={Journal of Computer Assisted Learning},
  volume={41},
  number={5},
  pages={e70103},
  year={2025},
  publisher={Wiley Online Library}
}

@article{breideband2025feasibility,
  title={A feasibility and implementation integrity study of the community Builder (CoBi): An AI-based collaboration support system in K-12 classrooms},
  author={Breideband, Thomas and Bush, Jeffrey B and Reitman, Jason G and Rose, Sierra and Weatherley, John and Penuel, William R and D’Mello, Sidney K},
  journal={International Journal of Artificial Intelligence in Education},
  pages={1--35},
  year={2025},
  publisher={Springer}
}

@article{zheng2025novel,
  title={A novel group cognitive graph approach for improving collaborative problem solving},
  author={Zheng, Lanqin and Shi, Zhe and Gao, Lei and Fan, Yunchao},
  journal={International Journal of Educational Technology in Higher Education},
  volume={22},
  number={1},
  pages={18},
  year={2025},
  publisher={Springer}
}

@article{slade2023collaborative,
  title={Collaborative team dynamics and scholarly outcomes of multidisciplinary research teams: A mixed-methods approach},
  author={Slade, Emily and Kern, Philip A and Kegebein, Robert L and Liu, Chang and Thompson, Joel C and Kelly, Thomas H and King, Victoria L and DiPaola, Robert S and Surratt, Hilary L},
  journal={Journal of Clinical and Translational Science},
  volume={7},
  number={1},
  pages={e59},
  year={2023},
  publisher={Cambridge University Press}
}

@article{george2023managing,
  title={Managing the strategic transformation of higher education through artificial intelligence},
  author={George, Babu and Wooden, Ontario},
  journal={Administrative Sciences},
  volume={13},
  number={9},
  pages={196},
  year={2023},
  publisher={MDPI}
}

@inproceedings{song2021survey,
  title={A survey of published literature on conversational artificial intelligence},
  author={Song, Xinmeng and Xiong, Ting},
  booktitle={2021 7th International conference on information management (ICIM)},
  pages={113--117},
  year={2021},
  organization={IEEE}
}

@article{liu2023incorporating,
  title={Incorporating a reflective thinking promoting mechanism into artificial intelligence-supported English writing environments},
  author={Liu, Chenchen and Hou, Jierui and Tu, Yun-Fang and Wang, Youmei and Hwang, Gwo-Jen},
  journal={Interactive Learning Environments},
  volume={31},
  number={9},
  pages={5614--5632},
  year={2023},
  publisher={Taylor \& Francis}
}

@inproceedings{zhao2026ontoscope,
  title={OntoScope: Using a Divergent-Convergent Interaction Framework to Support LLM-based Ontology Scoping},
  author={Zhao, Yihang and Mero{\~n}o Pe{\~n}uela, Albert and Simperl, Elena},
  booktitle={Proceedings of the 31st International Conference on Intelligent User Interfaces},
  pages={67--84},
  year={2026}
}

@inproceedings{zhao2025leveraging,
  title={Leveraging large language models for ontology requirements engineering},
  author={Zhao, Yihang},
  booktitle={European Semantic Web Conference},
  pages={254--264},
  year={2025},
  organization={Springer}
}

@article{zhaosurvey,
  title={A Survey on Interaction Design with Large Language Models for Ontology Requirements Elicitation with Competency Questions},
  author={Zhao, Yihang and Hu, Xi and Neate, Timothy and Pe{\~n}uela, Albert Mero{\~n}o and Simperl, Elena}
}

@inproceedings{zhao2024improving,
  title={Improving ontology requirements engineering with ontochat and participatory prompting},
  author={Zhao, Yihang and Zhang, Bohui and Hu, Xi and Ouyang, Shuyin and Kim, Jongmo and Jain, Nitisha and De Berardinis, Jacopo and Mero{\~n}o-Pe{\~n}uela, Albert and Simperl, Elena},
  booktitle={Proceedings of the AAAI Symposium Series},
  volume={4},
  number={1},
  pages={253--257},
  year={2024}
}

@article{zhao2024using,
  title={Using large language models for ontoclean-based ontology refinement},
  author={Zhao, Yihang and Vetter, Neil and Aryan, Kaveh},
  journal={arXiv preprint arXiv:2403.15864},
  year={2024}
}

@misc{zhao2026designguidanceaddressingoverreliance,
      title={Design Guidance Towards Addressing Over-Reliance on AI in Sensemaking}, 
      author={Yihang Zhao and Wenxin Zhang and Amy Rechkemmer and Albert Meroño Peñuela and Elena Simperl},
      year={2026},
      eprint={2603.08903},
      archivePrefix={arXiv},
      primaryClass={cs.HC},
      url={https://arxiv.org/abs/2603.08903}, 
}

@inproceedings{lisena2026data,
  title={Data-Driven Storytelling: Bridging Knowledge Graphs, GenAI, and Narrative (DDS 2026)},
  author={Lisena, Pasquale and Pellegrino, Maria Angela and Gan, Lisa-Yao and Zhao, Yihang and Xing, Yiwen},
  booktitle={The 25th International Semantic Web Conference (ISWC 2026)},
  year={2026},
  organization={Springer Nature}
}

@inproceedings{simperl2025automatic,
  title={Automatic Verification of References of Wikidata Statements},
  author={Simperl, Elena and Rodrigues, Odinaldo and Mero{\~n}o-Pe{\~n}uela, Albert and Amaral, Gabriel Maia Rocha and Gavenski, Nathan and Redi, Miriam and Zhang, Bohui and Zhao, Yihang and others},
  booktitle={Wikidata and Research},
  year={2025}
}

@inproceedings{simperl2025introducing,
  title={Introducing ProVe for Wikidata: Automatic Verification of References},
  author={Simperl, Elena and Rodrigues, Odinaldo and Penuela, Albert Merono and Amaral, Gabriel Maia Rocha and Gavenski, Nathan and Kim, Jongmo and Redi, Miriam and Zhao, Yihang},
  booktitle={12th Annual Wiki Workshop},
  year={2025},
  organization={Wikimedia Foundation}
}

@incollection{zhang2025trustworthy,
  title={Trustworthy knowledge graphs: Practices and approaches},
  author={Zhang, Bohui and Koutsiana, Elisavet and Zhao, Yihang and Mero{\~n}o-Pe{\~n}uela, Albert and Simperl, Elena},
  booktitle={Handbook on Neurosymbolic AI and Knowledge Graphs},
  pages={363--384},
  year={2025},
  publisher={IOS Press}
}

@inproceedings{zhao2024user,
  title={User experience in dataset search},
  author={Zhao, Yihang and Pe{\~n}uela, Albert Mero{\~n}o and Simperl, Elena},
  booktitle={International Conference on Computer-Human Interaction Research and Applications},
  pages={113--130},
  year={2024},
  organization={Springer}
}

@inproceedings{zhao2021discriminative,
  title={A Discriminative Deep Neural Network for COVID-19 Detection},
  author={Zhao, Yihang},
  booktitle={2021 International Conference on High Performance Big Data and Intelligent Systems (HPBD\&IS)},
  pages={65--68},
  year={2021},
  organization={IEEE}
}

@inproceedings{luo2021survey,
  title={A survey of transformer and GNN for aspect-based sentiment analysis},
  author={Luo, Wenqing and Zhang, Wei and Zhao, Yihang},
  booktitle={2021 International Conference on Computer Information Science and Artificial Intelligence (CISAI)},
  pages={353--357},
  year={2021},
  organization={IEEE}
}

@inproceedings{hu2025designing,
  title={Designing Interactions with Generative AI for Art and Creativity: A Systematic Review and Taxonomy},
  author={Hu, Xi and Xing, Yiwen and Cai, Xudong and Zhao, Yihang and Cook, Michael and Borgo, Rita and Neate, Timothy},
  booktitle={Proceedings of the 2025 ACM Designing Interactive Systems Conference},
  pages={1126--1155},
  year={2025}
}





\end{document}